\documentclass{article}
\usepackage[utf8]{inputenc}
\usepackage{mathtools,amsthm,amssymb,fullpage}
\usepackage{amsmath}
\usepackage{tikz-cd}
\usepackage{mathtools}
\usepackage{commath}
\usepackage{mathabx}
\usepackage[T1]{fontenc}
\usepackage{enumitem}
\usepackage{bbm}
\usepackage[backend=biber, style=nature]{biblatex}
\usepackage{authblk}
\usepackage{lineno}
\newcommand{\ket}[1]{| #1 \rangle}

\title{$BQP$ is not in $NP$}
\author[1]{Jonah Librande}
\affil[1]{University of Washington, Seattle, Washington 98195-1560, USA}

\begin{document}

\maketitle

\section{Introduction}
Quantum computers are often cited as superior to classical computers\cite{harrow}, and some actors have even claimed to have demonstrated their supremacy over their classical predecessors\cite{goog}, in the sense of exhibiting a computational problem to which the quantum machine produces a correct answer more efficiently than any classical machine can. Proving the existence of such an advantage would imply the existence of \textit{quantum} algorithms which solve problems we believe to be intractable on modern \textit{classical} machines. However, a single demonstration cannot prove the existence of such an advantage, as it could be the case that we are merely (and unknowingly) demonstrating the inadequacy of our own classical algorithms, when there does exist a more ingenious classical solution that reduces the performance gap. The proof of an advantage would require at least showing that $BQP$, the class of problems which can be computed efficiently by a quantum computer, is not contained in $P$, the class of problems which can be computed efficiently by a classical computer, implying the existence of problems within $BQP$ which cannot be done efficiently on a classical computer. Here, I provide a stronger proof, exhibiting a family of decision problems which are trivially contained in $BQP$, but which also cannot even lie within $NP$, the parent class of $P$. This has significant implications for any scientist who wishes to solve a computationally demanding problem using existing hardware: it demonstrates that uncountably many problems are exponential faster on quantum machines than on classical ones.

However, I must forecast that in this paper, I will work over a complete computational basis, which is to say, I permit the use of any possible one or two qubit gate in our circuits. This will require a bit of exposition to make our basis usable in a complexity-theoretic context.

\section{Working with a complete basis}
We describe some details of working over our basis. We need in particular to know how to interpret the "construction" of circuits in $BQP$ if the basis is so large, as the standard definitions generally assume that circuits in $BQP$ have descriptions which are generated by a Turing machine which operates over a finite alphabet of symbols. This section is a summary of material developed in another paper, which is to be submitted.

\subsection{U-strings defined}
To specify an element of our uncountable basis, I use the notion of a \textbf{U-string}. These are constructed by starting with the set $A$, which I define to be the union of the familiar alphanumeric alphabet of classical computation with the set of all possible 1 and 2 qubit gates. We consider each element of $A$ to be a character, and then define a U-string to be any finite tuple over $A$. This will allow us to generate reasonable-sized descriptions of arbitrary circuits over our basis. 

\subsection{The canonical encoding of a circuit}
For what follows, we would like a standard way of encoding a circuit as a U-string. For convenience, I will always assume that circuits come endowed with an ordering of their qubits; consequently, a reordering of the qubits in a circuit will generally produce a circuit considered to be distinct from the original circuit. 

To produce our encoding, we can start by dividing the circuit into "layers," or groups of gates which may act simultaneously, in the sense that all gates in a layer (1) act after all preceding layers, (2) act before all following layers, and (3) all the gates in the layer commute with one another \textit{and} share no arguments pairwise except for perhaps a qubit which is used only as a control qubit, by gates in that layer. For any gate in our basis, it is also in our alphabet $A$ and we can refer to it by the corresponding character $g \in A$. It acts on at most two qubits, of indices $i$ and $j$, where $i$ is the index of the first argument $g$ takes, and $j$ the index of the second. We encode this singular gate as $g[i, j]$. 

To encode a layer of gates, we concatenate the encodings of all gates in the layer, starting with the gate having the smallest index argument. If two gates both have the same smallest index argument, then this shared qubit must be used as a control by both, or else the two gates cannot be in the same layer. Each then acts on one more qubit aside from the control qubit (and only one more, since the basis consists of one and two qubit gates only), and we first encode the gate whose second argument is of a smaller index in this case. 

To encode the entire circuit, we simply concatenate the encodings of the layers, in their order of action, so that the first gate we see in the encoding, reading from left to right, will be the first gate (read: gate having the smallest index among the qubits it acts upon) of the first layer to act on the register. We concatenate to the front of this concatenation of layers, the integer number of qubits in the circuit.

As an example of this encoding, consider the circuit on two qubits which has a Hadamard gate $H$ on the first qubit, and a controlled not gate $c\sigma_x$ after that, with the first qubit acting as the control. The canonical encoding of this gate would be 
\begin{align*}
    2 H[0] c\sigma_x[0, 1],
\end{align*}
with the regular convention that an enumeration of a register starts at 0.

It is useful here to define our terminology for the phase gates: we define
\begin{align*}
    P \left (\theta \right) = 
    \begin{bmatrix}
         1 & 0\\
         0 & e^{i \theta}
    \end{bmatrix},
\end{align*}
and refer to the controlled variant of this gate by $cP(\theta)$. We refer to both as phase gates.

\section{The problem family $AQCE^*$}
We define a family of decision problems, each indexed by any finite number of gates from our basis. We shall call this family of problems $AQCE^*$, for "aliased quantum circuit evaluation." An example member of this family would be $AQCE(\sigma_x, H)$, where $\sigma_x$ is the quantum-NOT gate and $H$ is the Hadamard gate. We note that this class will be distinct from the problem class $AQCE(H, \sigma_x)$; the order of the parameters matters! We use $AQCE^*$ to refer to the family of all such problems, for all valid parameters, but do not consider $AQCE^*$ as a decision problem class in its own right. We shall delay the definition of the actual decision problem until the end of this section, as we must first describe the problem instances.

Every problem instance in this family is made from a quantum circuit satisfying a certain promise we shall soon specify, and to each such quantum circuit, we can locate a problem in the $AQCE^*$ family. We let $U = (u_1, u_2, \ldots, u_m)$ be some set of basis gates, and consider the class $AQCE(U)$. The instances of this class will be strings $s \in AQCE(U)$ (note that these are regular alphanumeric strings, not U-strings!) satisfying two sorts of requirements, several of form and one of substance. 

The requirements in terms of form are somewhat lengthy to describe, but not difficult to understand: we require that the first characters of $s$ be a number $n$ in decimal. After that, the rest of the string shall be composed of a regular series of substrings. Each substring will begin with an open parenthesis, "(", and conclude with a close square bracket, "]". These characters will occur nowhere within substrings, and in fact, delineate the start and end of substrings. Each substring will be of one of the following forms, with $k$ as an integer in the range $1$ to $m$ inclusive (with $m$ being the number of gates in $U$ as given above), $r$ an integer in the range $1$ to $n-1$ inclusive, and $i, j$ integers in the range $0$ to $n-1$ inclusive:
\begin{enumerate}
    \item $(k)[j]$;
    \item $(k)[i, j]$;
    \item $(*r)[j]$; and
    \item $(c*r)[i, j]$.
\end{enumerate}
We shall refer to a substring as of the type 1, 2, 3, or 4, referring to the numbered entries above. A given instance string $s$ may have any number of these substrings within it.

We may interpret any string $s \in AQCE(U) = AQCE(u_1, u_2, \ldots, u_m)$ as describing a quantum circuit. This is done by using the definition of the class as a dictionary to replace the substrings $(k)$ by gate characters from the U-string alphabet, in the following way:
\begin{enumerate}[label=(\alph*)]
    \item Whenever we see a substring of type 1 or 2, replace the characters "$(k)$" by the U-string character $u_k$. Well-formed instances will always be such that whenever the substring is of type 1, the gate $u_k$ is a one-qubit gate, and whenever it is of type 2, the gate is a two-qubit gate. 
    
    \item Whenever we see a substring of type 3, replace the symbol $(*r)$ by the gate $P( \pi / 2^r)$. 
    
    \item Whenever we see a substring of type 4, similarly, we replace the symbol $(c*r)$ by the gate $cP(\pi / 2^r)$.
\end{enumerate}
Observe that this will give exactly a canonical encoding of a circuit, assuming the input string was of the form described above. We shall call this process (of determining a quantum circuit from a string $s \in AQCE(U)$) \textbf{de-aliasing}. We observe that it depends critically on the problem class to which it belongs; this will become relevant later.

The requirement of substance on the string $s$ is really a requirement on the circuit it encodes, when we de-alias it using the parent class $AQCE(U)$: call this circuit $C_{U,s}$. We can initialize every qubit in the input register of $C_{U, s}$ to the state $\ket{0}$; assuming we do so, we may ask what the probability is of measuring $\ket{1}$ from the first qubit of the output register. Call this probability $p$. Our requirement of substance is that either $p \geq 2/3$ or $p \leq 1/3$. 

Any string which satisfies these two requirements for some class $AQCE(U)$ (which is to say, it satisfies the structure requirements and decodes into the proper sort of circuit when using the list of gates $U$ which come with the class $AQCE(U)$) is then a member of $AQCE(U)$. 

So, we have that $s \in AQCE(U)$ corresponds to a unique circuit. The question germane to $AQCE(U)$ is then this: \textit{when the circuit described by $s$ is run with its register initialized to the all-zero tensor state, is the probability that a measurement of the first qubit of the output will be $\ket{1}$ greater than 2/3, or is it less than 1/3?} Recall that we are promised one of these will hold. 

Of course, these instance strings are reasonable descriptions of quantum circuits, by any metric I can imagine. These strings are all perfectly exact: to each $s \in AQCE(U)$ corresponds a unique quantum circuit, which can be deduced in a very straightforward way. I struggle to think of a definition of a "description of a quantum circuit" which can exclude $ACQE^*$ instances while still allowing anything a regular Turing machine may be able to produce for finite precision, for the output of a Turing machine, over some finite alphabet, is merely a set of characters, and if we are to agree that the string $cX$ means "controlled-NOT" even though "$cX$" is clearly a pair of Latin characters and not a unitary having a certain matrix representation, or a piece of hardware, then I see no reason why we cannot agree in context that $(5)$ refers to the quantum gate $P(\pi^2)$. The string $(5)$ is as much an actual quantum gate as the string $cX$ is. 

I will then consider the string $s \in AQCE(U)$ for some $U$ to constitute a valid description of a quantum circuit, and so the problems $AQCE^*$ are contained in $BQP$. Their membership in the classical hierarchy is to be determined.

\subsection{Two example $AQCE^*$ instances}
For clarity, I would like to show two $AQCE^*$ instances, and the canonical encodings they represent. We consider the string $s = 2(1)[0](2)[0,1]$, which is not an $AQCE^*$ instance on its own; to be considered an instance, we must know the specific class to which it belongs. It becomes an instance when endowed with class membership, such as, say, $s \in AQCE(H, c\sigma_x)$, where $H$ is the Hadamard gate, and $c\sigma_x$ is the controlled NOT gate. We need the parameters of the class to de-alias the instance, and we can determine this $s$ considered as a member of $AQCE(H, c\sigma_x)$ represents the circuit $2 H[0] c\sigma_x[0, 1]$, as the reader can verify by using the de-aliasing procedure given above. 

The parameters of the class to which our string belongs determines the instance circuit entirely, as can be seen with the distinct decision problem given by $s \in AQCE(H, cH)$, where $cH$ is the controlled Hadamard gate; this problem has an instance circuit $2 H[0] cH[0, 1]$. I wish to stress that this shows the string $s$ does not in general determine a unique circuit, and this same string will, depending on the $AQCE^*$ class to which it is considered as a member of, refer to a multiplicity of different circuits.

\section{The proof of noncontainment}
We at last are in a position to prove the main result. 

\textbf{Theorem.} \textit{$BQP \nsubset NP$.}

\begin{proof}
Suppose for contradiction that $BQP \subset NP$. Then, since every $AQCE^*$ problem is contained in $BQP$, they will also also lie in $NP$. This means that for every member class $AQCE(U)$ of the family $AQCE^*$, there will exist a nondeterministic Turing machine $M_U$ which decides every $x \in AQCE(U)$. 

We then let $H = \{1/2^j: j \in \mathbb{Z}^+\}$, and pick $x, y \in (0, 2\pi) \setminus H$ such that $x \neq y$. Since these two reals are distinct, their binary expansions must differ at some finite index, and for concreteness, we say they first differ at the $k$-th bit of their expansions. 

These two reals define distinct phase gates $P(x)$ and $P(y)$, and these phase gates both share the eigenstate $\ket{1}$, though they each associate a different eigenvalue to it -- we have $P(x) \ket{1} = e^{ix} \ket{1}$ and $P(y) \ket{1} = e^{iy} \ket{1}$. Now, we would like to build phase estimation circuits to estimate the phase of each of these $\ket{1}$ eigenvalues, to which we will then append a few auxiliary gates; call the circuit we wish to construct for $P(x)$ by $C_x$, and the circuit for $P(y)$ by $C_y$. We note here that the preparation of the eigenvalue of the phase gates is trivial -- assuming every qubit of the register starts in the state $\ket{0}$, we need only place a single NOT gate on the qubit which we wish to put in the eigenstate. In this way, we can absorb the preparation of the eigenstate into the circuit.

As noted in the appendices, we can enlarge a phase estimation circuit by an extra, finite number of qubits to increase the probability of receiving an accurate measurement to any finite level we desire\cite{errors}. In particular, we need only append some finite number $r$ of qubits to one of the phase estimation circuits to obtain an accurate estimate of the first $k$ bits of the phase with probability at least 2/3, say, 0.8. We will then choose the register size of both circuits to be the same number, such that both circuits will have probability at least 0.8 of measuring an accurate estimate of the first $k$ bits of the phases.

The phases being estimated are in fact the reals $x$ and $y$, and since these values first differ at the $k$-th bit, then the stipulation on measurement probabilities implies that, without loss of generality, the $k$-th qubit of the output in the circuit for $P(x)$ will be measured to be in the state $\ket{0}$ with probability at least 0.8, while the same qubit of the output in the circuit for $P(y)$ will be measured to be in the state $\ket{1}$ with probability at least 0.8.

To conclude these two circuits, we append at the end a qubit swap operation, between the $k$-th qubit and the first qubit in the register. This can be accomplished by a trio of controlled-NOT gates, which we shall refer to as $cX$ gates, to distinguish them from regular NOT ($X$) gates. This means that the \textit{first} qubit in the register of the output of $C_x$ will, with probability at least 0.8, be measured to be in the state $\ket{0}$, while the same qubit for $C_y$ will be measured to be in the state $\ket{1}$ with probability at least 0.8. These probabilities will be of importance shortly.

Now, we consider encoding these two circuits as problem instances in the family $AQCE^*$. We observe that, aside from $W$ gates, each circuit is comprised of just four gates: the Hadamard gate $H$, the NOT gate $X$ (used to realize the eigenstate $\ket{1}$), the corresponding controlled phase gate $cP(x)$ or $cP(y)$ (depending on the circuit in question), and the controlled-NOT gate $cX$ (for the qubit swap). In addition, we may consider the canonical U-string encodings of both circuits; call the encodings for $C_x$ and $C_y$ by $u_x$ and $u_y$, respectively. These encodings will be the same length, and in fact, they will not match at their $j$-th characters if and only if the character is a controlled phase gate, a fact which follows from the procedure to create the phase estimation circuit, since we fixed both circuits to estimate the phase to the same precision (which is to say, we fixed them both to have the same register size) and since both gates of interest share the same eigenstate of interest; the procedure is agnostic to the action of any particular gate, only requiring that we have access to the controlled variant of the gate, and that the eigenstate of interest is somehow prepared in the lower register. Of course, for our procedure, we can easily prepare the eigenstate of interest -- it is just $\ket{1}$, for both circuits, which is easily prepared. 

The similarity of these encodings means that there is a string $s$ such that when considered as a problem instance $s \in AQCE(H, X, cP(x), cX)$, the circuit in instance $s$ is the circuit $C_x$, and while considered as an instance $s \in AQCE(H, X, cP(y), cX)$, the circuit in the instance is $C_y$. This because the circuit we construct from such an instance is decided both by the form of the instance string, and the dictionary of gates which defines the problem class. The string is the same for both problems, and the dictionaries only differ in the third gate, which is also the only point at which the circuits $C_x$ and $C_y$ differ. Thus, de-aliasing the instances will give these two circuits, from the same string, using different dictionaries.

Since we assume $BQP \subset NP$, there exists a nondeterministic Turing machine $M_x$ which decides every instance of $AQCE(H, X, cP(x), cX)$, meaning that when fed an instance $p \in AQCE(H, X, cP(x), cX)$, $M_p$ will accept $p$ if, after being run on a register initialized to the all-$\ket{0}$ tensor state, a measurement of the first qubit of the output of the circuit represented by $p$ will return $\ket{1}$ with probability at least 2/3, and reject it if the probability is less than 1/3. Similarly, there exists an analogous nondeterministic Turing machine $M_y$ for $AQCE(H, X, cP(y), cX)$ with identical acceptance criteria. 

I claim that the machines $M_x$ and $M_y$ must be distinct. To see this, we note that $M_x$ must reject $s \in AQCE(H, X, cP(x), cX)$, as the circuit this instance corresponds to ($C_x$) has an output whose first qubit will \textit{not} be measured in the state $\ket{1}$ with probability at least 0.8. However, $M_y$ must accept $s \in AQCE(H, X, cP(y), cX)$, as the circuit this instance corresponds to ($C_y$) will have an output whose first qubit will be measured to be in the state $\ket{1}$ with probability at least 0.8. Of course, the instance string itself $s$ is the same in both cases, and a single Turing machine (even if nondeterministic!) cannot both reject and accept the same string, so the two Turing machines must be distinct.

The values $x, y \in (0, 2\pi) \setminus H$ were arbitrary, so that for any distinct $a, b \in (0, 2\pi)\setminus H$, the nondeterministic Turing machines $M_a$ and $M_b$ deciding the problem classes $AQCE(H, X, cP(a), cX)$ and $AQCE(H, X, cP(b), cX)$ must be distinct. However, since $H$ is countable, the set $(0, 2\pi) \setminus H$ is uncountable, and we thus require an uncountable number of nondeterministic Turing machines to decide the classes $AQCE(H, X, cP(a), cX)$ for each $a \in (0, 2\pi) \setminus H$ -- but this is impossible, as the set of all nondeterministic Turing machines is countable! Thus, we have a contradiction, and our original assumption that $BQP \subset NP$ must fail to hold.

\end{proof}

\section{Acknowledgement}
I would like to thank Kenneth Roche for his help in preparing this manuscript.

\section*{Appendices}
Collected here are bit of information that may be useful to the reader, but which has no easy place ot be inserted into the body of the main article.

\subsection*{Appendix A: $BQP$ defined}
There are multiple definitions of $BQP$ floating around. Here are a representative pair as concerns regular quantum computation over a finite basis:
\begin{enumerate}
    \item (Listing gates via Turing machine\cite{arora}.) A boolean function $f: \{0, 1\}^* \to \{0, 1\}$ is in $BQP$ if there is some polynomial $p$ such that for every $x \in \{0, 1\}^n$, there exists a quantum circuit $C_{f(x)}$ such that:
    \begin{enumerate}
        \item $C_{f(x)}$ has no more than $n + p(n)$ qubits in its register;
        \item There are no more than $p(n)$ elementary gates in $C_{f(x)}$, and there exists some polynomial-time Turing machine that when fed $1^n$, outputs the elementary gates to be applied.
        \item A measurement of the first qubit of $C_{f(x)}$ will be equal to $f(x)$ with probability at least 2/3.
    \end{enumerate}
    
    \item (Families of descriptions of quantum circuits\cite{kitaev}.) A boolean function $f$ is in $BQP$ if there is a classical algorithm computing a function of the form $x \mapsto Z(x)$ in polynomial time, where $Z(x)$ is a description of a quantum circuit which computes $f(x)$ on empty input.  
\end{enumerate}
Both of these definitions require that we obtain a descriptive listing of gates from some Turing machine in polynomial time. However, as Turing machines by definition only have access to a finite alphabet, they cannot describe an arbitrary element of the uncountable basis we are using, and thus it appears that we are vastly restricting the power of quantum computation by using this Turing machine-based conception of $BQP$. This tension is at least bandaged, if not resolved, in the $AQCE^*$ problem classes, which allow the exact description of arbitrary circuits using the alphabet of classical computation.

\subsection*{Appendix B: The quantum phase estimation routine}
We gather here a few facts about the quantum phase estimation routine which are relevant in the main text; most of the general information can be found in \cite{nichu}. The routine involves two registers, which I shall call the "main" register and "phase" register; the main register will be initialized to the state $\ket{0}^{\otimes n}$, where $n$ is determined by the desired accuracy of the estimation to be obtained, while the phase register will be prepared in a state determined by the gate which we want to estimate.

In general, for some gate $U$ and a eigenstate $\ket{\psi}$ of $U$ prepared in the phase register, we will have $U \ket{\psi} = e^{i \theta} \ket{\psi}$ for $\theta \in [0, 2\pi]$. The quantum phase routine, in its conventional form, allows us to construct an estimate for $\theta$, given some number of copies of a controlled version of some gate realizing $U$, which we call $cU$.

The circuit can be constructed, aside from the prepared state $\ket{\psi}$ and the copies of $cU$, using only two sorts of gates:
\begin{enumerate}
    \item Hadamard gates, and
    \item The set of gates $cW(k)$ for $k$ ranging from 1 to $n-1$, where $n$ is the number of qubits in the main register.
\end{enumerate}
Together with the controlled gates $cU$, this comprises all that is needed to run the circuit. Once the circuit is constructed and run, some subset of the main register is measured. The estimate for the phase is obtained by interpreting the value measured as being the value $2^k \theta$, where $k$ is the number of qubits we are measuring.

There are two pertinent facts about the phase estimation routine I wish to draw attention to: first, if we wish our measurement of the main register to give an accurate estimate in the first $k$ qubits with probability at least $1 - \epsilon$ for any $\epsilon > 0$, then this can be done by adding a finite, $\epsilon$-dependent number of qubits to the main register \cite{errors}. Second, the phase estimation circuit on $n$ qubits will have a main register which is practically independent of the controlled unitary $cU$ that we wish to estimate the phase of: if two unitaries $U$ and $V$ have the same eigenstate, then creating a phase estimation circuit with $n$ qubits (in the main register) for the unitary $U$, and then replacing every gate $cU$ by the gate $cV$, with the same control and acted upon qubits, will exactly turn the circuit into a phase estimation circuit on $n$ qubits (in the main register) for $V$.

\subsection*{Appendix C: From circuits to $AQCE^*$}
Let $g$ be a mapping which takes as a domain all quantum circuits $C$ which are such that, when run on a register initialized to the state $\ket{0}^{\otimes n}$the probability of measuring the first qubit of the output to be in the state $\ket{1}$ is either greater than 2/3 or less than 1/3. Fed such an input, the function $g$ outputs a pair $(s, AQCE(U))$, where $s$ is just some regular classical string over the finite alphabet of classical computing, and $AQCE(U)$ is a member of the family $AQCE^*$ for some list of gates $U$. This output tells us that the string $s$ is to be understood as an instance of the problem class $AQCE(U)$.

For any circuit $C$ on $n$ qubits, the action of $g$ proceeds in steps. First, it will canonically encode $C$ as a U-string. Then, we consider the set of all basis gates that occur in $C$; this set is necessarily finite, and the canonical encoding of $C$ gives these gates a natural enumeration. Of these gates, we consider all gates which are \textbf{not} of the form
\begin{align*}
    cP \left (\frac{\pi}{2^{k}} \right) = 
    \begin{bmatrix}
         1 & 0 & 0 & 0\\
         0 & 1 & 0 & 0\\
         0 & 0 & 1 & 0\\
         0 & 0 & 0 & e^{i \frac{\pi}{2^k}}
    \end{bmatrix}
\end{align*}
or
\begin{align*}
    P \left (\frac{\pi}{2^{k}} \right) = 
    \begin{bmatrix}
         1 & 0\\
         0 & e^{i \frac{\pi}{2^k}}
    \end{bmatrix}
\end{align*}
for integer $k \in [1, n-1]$; gates of these two forms will be referred to, in shorthand, as $W$ gates. They carry an integer index which refers to the power of $2$ in the denominator of the phase index; as an example, the former of the two preceding gates will be referred to as $cW(k)$ while the latter will be just $W(k)$. The remaining, non-$W$ gates in $C$ can then be enumerated, since the circuit is finite: let $h_1$ be the first unique basis gate that appears in the encoding of $C$ which is not of the above form, and then let $h_2$ be the second, and so on, until we have a set $\{h_j\}$ of all such basis gates appearing in $C$. Our $g$ will then perform an "aliasing" step, where it replaces every character from the gate set $B$ with a character from the alphabet of classical computation, before outputting the resultant, aliased, fully classical string into the class $AQCE(h_1, h_2, \ldots) = AQCE(\{h_j\})$; the second element of our output pair will be the class $AQCE(\{h_j\})$.

The aliasing step is a simple replacement routine. Consider a gate character $u$ in the encoding of $C$. Either $u$ is a $W$ gate or not. If it is not a $W$ gate, then $u$ is among the enumerated gates described above, and we may suppose it to be gate $h_i$. Then, $g$ will replace $u$ by the symbols $(i)$, where $i$ will be given in, say, decimal, using the characters of classical computation. 

If instead $u$ is a $W$ gate, then it is either of the form $cW(k)$ or it is of the form $W(k)$. If it is of the form $cW(k)$, then $g$ will replace $u$ by the characters $(c*k)$, where $c$ is just the regular alphabet character $c$ (or, equivalently, some sensible encoding thereof), $*$ is the star character, and $k$ is represented in decimal. If instead $u$ is of the form $W(k)$, we encode more simply as $(*k)$. 

Thus, for all gates in the uncountable set $B$, $g$ will replace them by a finite number of characters from the alphabet of classical computing. The resultant, fully classical string is then, as mentioned above, mapped into $AQCE(\{h_j\})$ as a realized instance. Every problem in any of the $AQCE^*$ can easily be seen to be obtainable in this way as the image under $g$ of some quantum circuit.

\printbibliography

@article{errors,
    author = "James M. Chappell and Max A. Lohe and Lorenz von Smekal and Azhar Iqbal and Derek Abbott",
    title = "A Precise Error Bound for Quantum Phase Estimation",
    journal = "PLoS ONE",
    volume = "6",
    number = "5",
    year = "2011",
    doi = "https://doi.org/10.1371/journal.pone.0019663"
}

@article{goog,
    author = "Frank Arute and \textit{et al.}",
    title = "Quantum supremacy using a programmable superconducting processor",
    journal = "Nature",
    volume = "574",
    year = "2019",
    pages = "505--510",
    doi = "https://doi.org/10.1038/s41586-019-1666-5"
}

@article{harrow,
    author = "Aram Harrow and Ashley Montanaro",
    title = "Quantum computational supremacy",
    journal = "Nature",
    volume = "549",
    year = "2017",
    pages = "203--209",
    doi = "https://doi.org/10.1038/nature23458"
}

@book{nichu,
    author = {Nielsen, Michael A. and Chuang, Isaac L.},
    title = {Quantum Computation and Quantum Information: 10th Anniversary Edition},
    year = {2011},
    isbn = {1107002176},
    publisher = {Cambridge University Press},
    address = {USA},
    edition = {10th}
}

@book{kitaev,
    author = "Mikhail N. Vyalyi and Alexander Shen and Alexei Kitaev",
    title = "Classical and Quantum Computation",
    year = "2002",
    publisher = "American Mathematical Society",
    series = "Graduate Studies in Mathematics"
}

@book{arora,
    author = "Sanjeev Arora and Boaz Barak",
    title = "Computational Complexity: A Modern Approach",
    year = "2009",
    publisher = "Cambridge University Press"
}

\end{document}